\documentclass[%
aps,prl,twocolumn,showpacs,superscriptaddress,groupedaddress
 reprint,
 amsmath,amssymb,
 aps,
]{revtex4-1}
\usepackage{xcolor}
\usepackage{graphicx}
\usepackage{dcolumn}
\usepackage{bm}

\begin{document}

\preprint{APS/123-QED}

\title{Probing the Lattice Anharmonicity of Superconducting\\ YBa$_2$Cu$_3$O$_{7-\delta}$ Via Phonon Harmonics }


\author{Alberto Ramos-Alvarez}%
 \affiliation{%
 ICFO-Institut de Ciencies Fotoniques\\
The Barcelona Institute of Science and Technology, 08860 Castelldefels, Barcelona, Spain
}%

\author{Nina Fleischmann}%

\affiliation{%
 ICFO-Institut de Ciencies Fotoniques\\
The Barcelona Institute of Science and Technology, 08860 Castelldefels, Barcelona, Spain 
}%

\author{Luciana Vidas}%

\affiliation{%
 ICFO-Institut de Ciencies Fotoniques\\
The Barcelona Institute of Science and Technology, 08860 Castelldefels, Barcelona, Spain 
}%

\author{Alejandro Fernandez-Rodriguez}%

\affiliation{%
 ICMAB-CSIC – Institut de Ciencia de Materials de Barcelona\\
 Campus de la UAB, 08193 Bellaterra, Catalonia, Spain 
}%

\author{Anna Palau}%

\affiliation{%
 ICMAB-CSIC – Institut de Ciencia de Materials de Barcelona\\
 Campus de la UAB, 08193 Bellaterra, Catalonia, Spain 
}%

\author{Simon Wall}%
	\email{simon.wall@icfo.eu}

\affiliation{%
 ICFO-Institut de Ciencies Fotoniques\\
The Barcelona Institute of Science and Technology, 08860 Castelldefels, Barcelona, Spain 
}%

\begin{abstract}
We examine coherent phonons in a strongly driven sample of optimally-doped high temperature superconductor YBa$_2$Cu$_3$O$_{7-\delta}$. We observe a non-linear lattice response of the 4.5\,THz copper-oxygen vibrational mode at high excitation densities, evidenced by the observation of the phonon third harmonic and indicating the mode is strongly anharmonic. In addition, we observe how high-amplitude phonon vibrations modify the position of the electronic charge transfer resonance. Both of these results have important implications for possible phonon-driven non-equilibrium superconductivity.   

\end{abstract}

\maketitle

\section{Introduction}

The physics of the copper-oxygen bond plays an important role in high-temperature cuprate superconductors. The bond controls both the degree of charge transfer between copper and oxygen ions and the spectrum of the lattice vibrations. Dynamical mean-field theory calculations have revealed that the copper-oxygen charge-transfer energy correlates strongly with the superconducting transition temperature~\cite{Weber} and the resonance is controlled by the apical oxygen bond strength~\cite{Kim}. Furthermore, transient control of copper-oxygen bond has been used to induce a superconducting-like state in a range of cuprate superconductors~\cite{Fausti_1, Hu, Mankowsky}, with  YBa$_2$Cu$_3$O$_{6.45}$ in particular, showing evidence for superconducting transport above room temperature~\cite{Kaiser}. 

While the claim of transient superconductivity is still debated~\cite{Orenstein, Chiriaco, Zhang}, one of the leading explanations for how such superconducting states could arise is based on phonon-phonon coupling~\cite{Mankowsky, Fechner}, similar to the process proposed for light-induced transitions in the manganites~\cite{Forst}. In this process an IR-active phonon, which can be resonantly driven by a laser field, is used to displace a Raman active phonon mode, which is responsible for the physical change. However, both experimental evidence for how phonon couple and how Raman active phonons can control the properties of high temperature superconductors are lacking. In this paper, we tackle both points in an optimally doped YBa$_2$Cu$_3$O$_{7-\delta}$ (YBCO) sample. First, we show that the 4.5 THz copper-oxygen bond has a strong anharmonicity at below the critical temperature for superconductivity, $T_c$, evidenced by the observation of phonon harmonics. As phonon-phonon coupling requires a strong lattice anharmonicity, our measurements provide new evidence that such coupling is possible. Secondly, we examine how perturbing the 4.5\,THz copper-oxygen bond can influence superconductivity, by spectrally-resolving the influence of the phonon contribution to the transient electronic structure. We find that the 4.5\,THz mode is able to shift the copper-oxygen charge transfer resonance energy, which is known to correlate with $T_c$ in equilibrium. These results provide new insights into how phonons can be controlled and be used to manipulate superconductivity.

\section{Lattice anharmonicity}
Infrared (IR) and Raman active phonons are the normal modes of the system and in the harmonic approximation, they are not coupled. In order to transfer energy between modes, some anharmonicity in the lattice potential is required. When an anharmonic oscillator is strongly driven, it produces oscillations both at its fundamental frequency and at higher harmonics, thus harmonics of the fundamental frequency of the Raman active mode can be used as a probe of the lattice anharmonicity. Recently observation harmonics emitted from a strongly and resonantly driven IR-active mode has enabled the mapping of the anharmonic lattice potential in a non-centrosymmetric material~\cite{Hoegen}. Here, we are interested in anharmonicity of a Raman active mode in a centrosymmetric material, which requires a different excitation and detection mechanisms in order to look for harmonics. The anharmonic lattice potential for a centrosymmetric material can be expressed as $V(x) = x^2 + \delta x^4$, where $x$ is the mode displacement and $\delta$ marks the degree of anharmonicity. In this potential, the time-dependent displacement of the ions is given by $x \propto \sin wt + \beta \sin 3wt$, where $\beta$ is the amplitude of the third harmonic signal and 

\begin{equation}
\label{eq1}
w^2 = (1 + \frac{3}{4} \delta x_0^2) w_0^2.
\end{equation}

Here $w_0$ and $x_0$ are the harmonic frequency and the initial displacement of the vibrational mode respectively. The amplitude of the third harmonic scales with the third power of the initial displacement~\cite{Banerjee}. Anharmonicity, therefore, gives rise to two key identifies, the appearance of third harmonic signal, which scales non-linearly with displacement, and a renormalization of the fundamental frequency. As Raman active modes cannot be linearly excited by electric dipole radiation, we excite the mode through a prompt displacive mechanism which results from the ultrafast generation of excited carriers, and probe the coherent lattice response through the modification of the optical properties. 

\section{Optical Measurements}

\begin{figure}
\centering

\includegraphics[scale=.95]{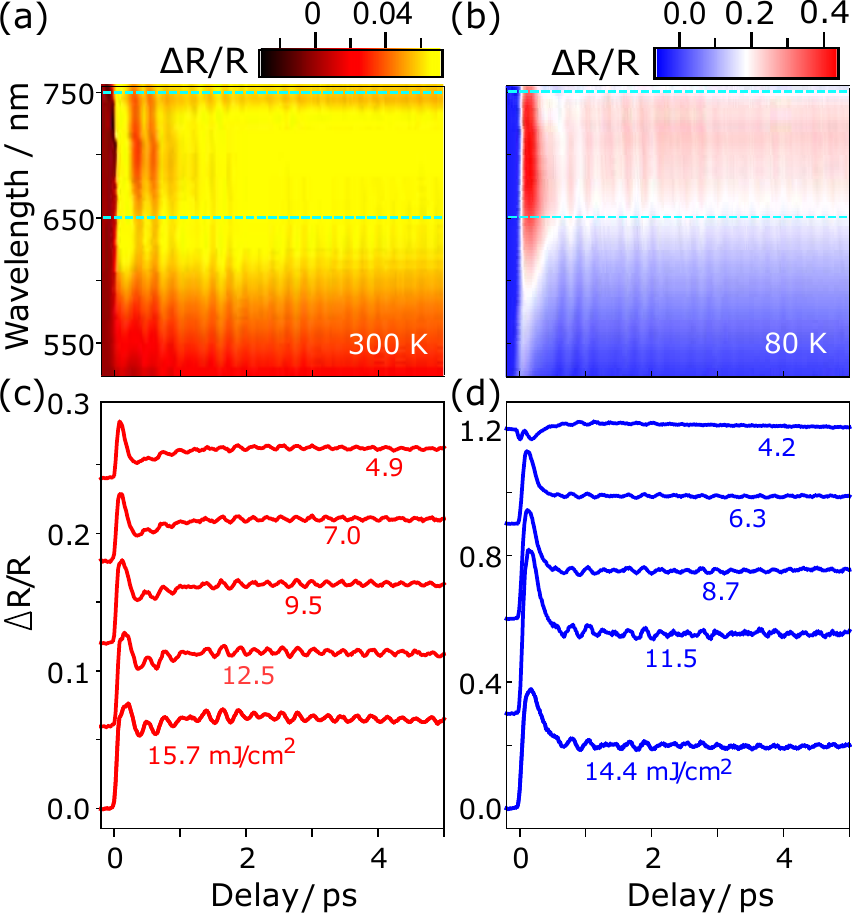}
\caption{\label{fig1} Broadband transient reflectivity of YBCO. Wavelength resolved signal at (a) 300 and at (b) 80 K, at the maximum pump fluence. Integration of $\Delta$R/R performed for several probe wavelengths, from 650 up to 750 nm is present in (c) and (d), for different pump fluences (mJ/cm$^2$), an offset is added to the signals to clarify.}

\end{figure}

A 250-nm-thick sample of epitaxial (001) oriented YBCO was grown by chemical solution deposition on a 5mm x 5mm (001) LaAlO$_3$ substrate. The critical temperature found at 90K was inductively measured with a SQUID magnetometer. The thin film was excited with 30 fs, 800 nm laser pulses at 5 kHz and was probed by broadband white-light generated in a sapphire crystal, which was dispersed onto a spectrometer. Measurements of the relative change of the reflectivity, $\Delta R/R$, were performed in a near-normal geometry, with the angle of incidence less than 10 degrees. We report measurements at different pump fluences with orders of magnitude higher than previous studies from $\sim$1 mJ/cm$^2$ up to 15 mJ/cm$^2$ at 80 K and at room temperature. While these fluences are more than sufficient to destroy superconductivity on the ultrafast timescale~\cite{Kusar, Giannetti, Coslovich, Rameau}, and thus will not lead to a light-induced superconducting state, we can still study how the phonons are generated and modify the properties of the material. Fig.~\ref{fig1} shows typical wavelength-resolved transient reflectivity spectra measured at room temperature (a) and at 80 K (b).

Although the amplitude varies, coherent phonons can be clearly seen across the whole spectral region, together with dynamics which are associated with the response of the photo-excited carriers~\cite{Peli}. The wavelength-integrated response, between 650 and 750\,nm, is shown in Fig.~\ref{fig1}(c) and (d) for several pump fluences. At 300\,K (Fig.~\ref{fig1}(c)) the dynamics show monotonic increment in amplitude with increasing pump fluence. The response is markedly different at 80\,K (Fig.~\ref{fig1}(d)). At low pump fluences the initial response is negative, but becomes positive signal at higher fluence. This sign change has been associated with the ultrafast transient melting of the superconducting condensate~\cite{Kusar, Giannetti, Coslovich}. While the incoherent dynamics of the charges is interesting in their own right~\cite{Peli}, here we focus on the dynamics of coherent phonons and how they influence the optical properties. For both temperatures, the phonon amplitude increases with increasing fluence and a clear beating can be observed, indicating the presence of two vibrational modes. 
\section{Lattice dynamics}

\begin{figure}
\centering

\includegraphics[scale=.9]{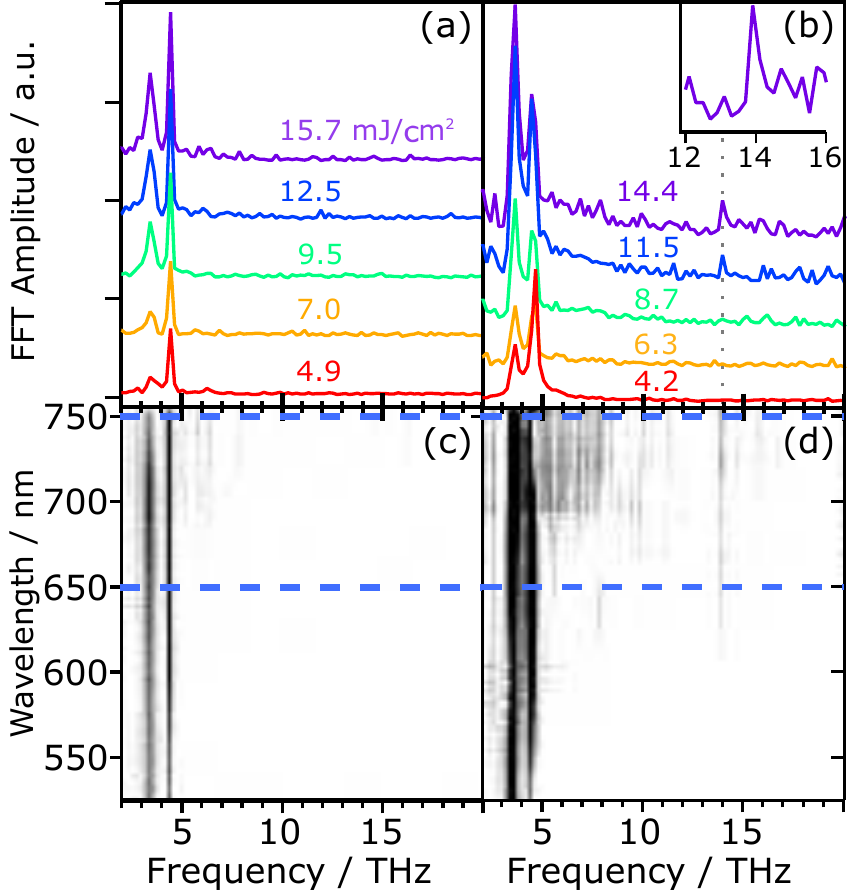}
\caption{\label{fig2} Fast Fourier Transform amplitude of the transient reflectivity signal after subtracting the electronic background contribution. Different pump fluences have been plotted with an offset added to clarify, from $\sim$ 4.5 up to 15 mJ/cm$^2$ at 300 K (a) and at 80 K (b). The signals were calculated by adding the contributions of a broadband of wavelengths from the probe light, from 650 up to 750 nm. Inset in (b) is a zoom to show the third harmonic of the Cu vibration mode at 4.5\,THz. Broadband resolved of the FFT amplitude is shown at 300 K (c) and at 80 K (d), dashed line mark the integration range of wavelengths.}

\end{figure}

In order to analyze the phonon amplitude in more detail, we removed the electronic response by fitting a wavelength dependent background. We then perform the fast Fourier transform (FFT) of the oscillations for each probe wavelength. The absolute value of the spectrally-integrated FFT, for different excitation powers, are shown in Fig.~\ref{fig2} at 300 K (a) and at 80 K (b). Figs. \ref{fig2}(c,d) show the wavelength-resolved responses at the highest fluence applied. In both temperatures the response is dominated by two phonon modes, one at $\sim$ 3.6\,THz and a second at $\sim$ 4.5\,THz. Conventional Raman scattering has shown that these vibrational modes have $A_g$ symmetry, and were identified as oscillations of the barium and the copper ions respectively \cite{Limonov}. These observations are in contrast to previous recent work, in which the barium mode was only observed at low temperatures~\cite{Novelli}, but is consistent with Raman~\cite{Limonov} and older time-resolved work~\cite{Albrecht}. 

At low fluences the $\sim$ 4.5\,THz copper mode dominates the signal for both temperatures. As the fluence is increased, both modes grow in intensity, but at different rates, and at 80\,K, the barium mode dominates at high fluences. More interestingly, we also observe a new peak emerging at $\sim$ 13.9\,THz for the two highest pump intensities applied in our experiment, 11.5 and 14.4\,mJ/cm$^2$. A mode at this frequency does not appear in the reported equilibrium Raman scattering spectra, but it is in excellent agreement with the third harmonic of the 4.5\,THz copper mode, marked by a dashed line in Fig.~\ref{fig2}(b), and is highlighted in the inset. 

\begin{figure}
    \centering
    \includegraphics{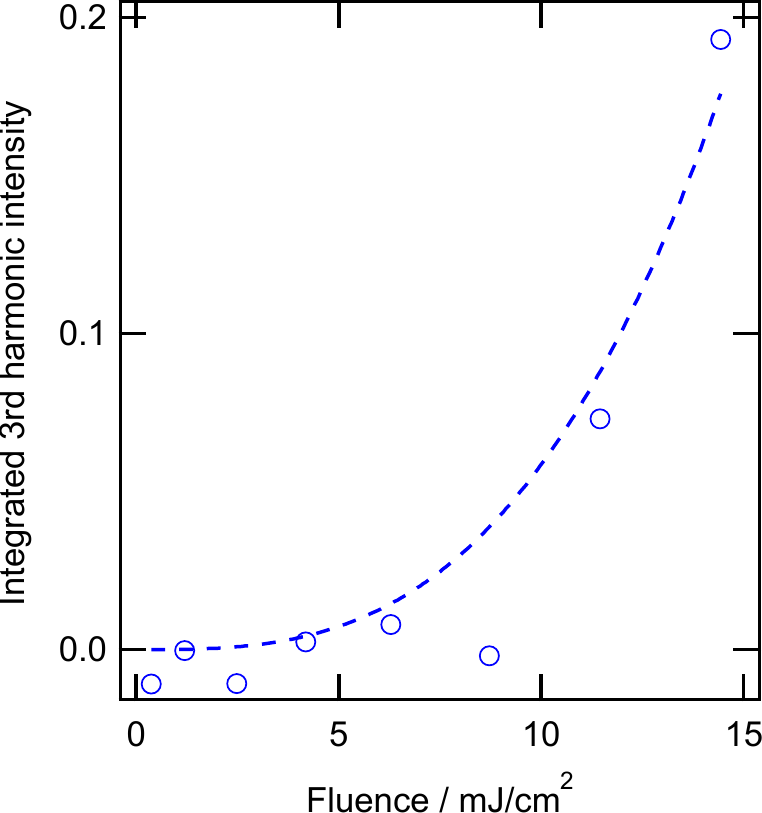}
    \caption{Fluence dependence of the third harmonic signal at 13.9 THz after background subtraction (blue circles) and cubic fit (dashed line). See the supplementary information for details of the background subtraction}
    \label{fig:fluence3rd}
\end{figure}

Figure~\ref{fig:fluence3rd} shows that the fluence ($F$) dependence of this vibrational mode scales as the $F^3$, consistent with the mode arising non-linearly from the anharmonicity of the lattice potential. We note that this mechanism is fundamentally different from previous observation of harmonics of the phonon frequency, which resulted from a non-linear dependence of the optical properties on the linear phonon response~\cite{Hase}. Ref~\cite{Hase} observed both even and odd harmonics at much lower excitation fluences due to the fact that the reflectivity change was non-linear with the phonon displacement, but the structure of the material was still physically oscillating at the fundamental frequency. Here, we claim the opposite. The reflectivity is responding linearly with the phonon displacement, but the phonon mode is reacting non-linearly due to anharmonicity. We exclude the non-linear optical effect as this can produce both even and odd harmonics, whereas we only observe the third harmonic, which is consistent with the site symmetry of the copper ion in the cuprates crystal structure. In principle, anharmonicity can generate higher odd harmonics, but these would lie outside of our current detectable limit of $\sim$ 20\,THz due to our finite time resolution. 

\begin{figure}
\centering

\includegraphics{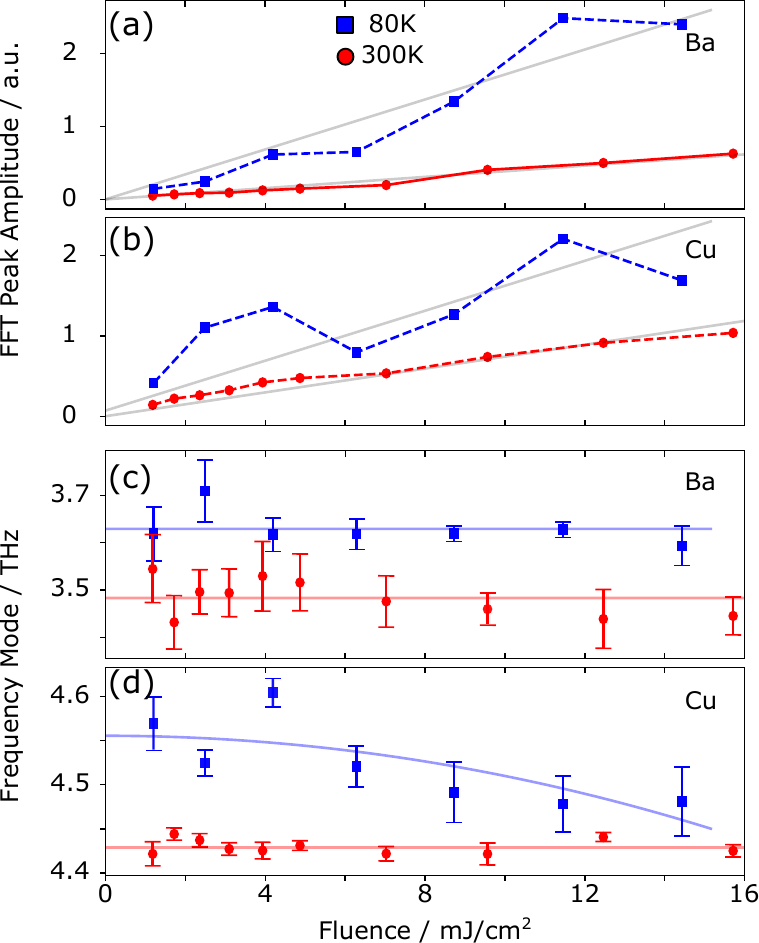}
\caption{\label{power_fun} Dependence on the amplitude and frequency of the two found coherent phonons with the pump fluence. The FFT peak amplitude is plotted at 300\,K (circles), and at 80K (squares), for the barium (a) and for the cooper mode (b). Grey solid lines are the linear fits of the experimental data. Central frequency of every mode versus pump intensity is shown for Ba (c) and for Cu (d) for the two temperatures studied. Solid lines follow the behavior of the data according to Eq.~\ref{eq1}.}
\end{figure}

In Fig.~\ref{power_fun} we show the pump-power dependence of the amplitude and fundamental frequency of each mode. The amplitude of the barium mode (Fig.~\ref{power_fun}(a)) shows linear dependence with laser fluence at room temperature, and a slight non-linearity at 80 K. The induced change in the optical properties is approximately four times larger at 80\,K for equivalent fluences. The copper mode (Fig.~\ref{power_fun}(b)), shows roughly the same level of increase in amplitude on cooling, but while the room temperature data has a linear behavior with laser fluence, the response at 80\,K is clearly non-monotonic. Figs \ref{power_fun}(c) and (d) show how the barium and copper vibrational frequency depends on fluence, respectively. At room temperature, the frequency of both the barium and copper modes was found to be independent of the pump intensity. At 80 K, both vibrational modes present a considerable blue shift, compared to the room temperature data, consistent with observations in Raman scattering measurements \cite{Limonov}. At low temperatures, the barium mode frequency is still independent of fluence, indicating both that there is no average heating effect in the sample, and that the mode does not soften with photoexcitation. In contrast, the copper vibrational mode shows a significant red-shift when the fluence is increased, approaching the room temperature values at the highest fluences.

Eq. \ref{eq1} can be used to fit the fluence dependence for the copper mode frequency at 80\,K, shown Fig.~\ref{power_fun}(d), by assuming that the initial displacement is linearly dependent on the excitation pump fluence. While we cannot exclude photo-induced dephasing as the origin of the frequency shift, the frequency shift, together with the third harmonic and its non-linear scaling with fluence provides strong evidence supporting an anharmonic Cu vibration mode. 

\section{Phonon modulation of the charge transfer resonance}

\begin{figure}
\centering
\includegraphics{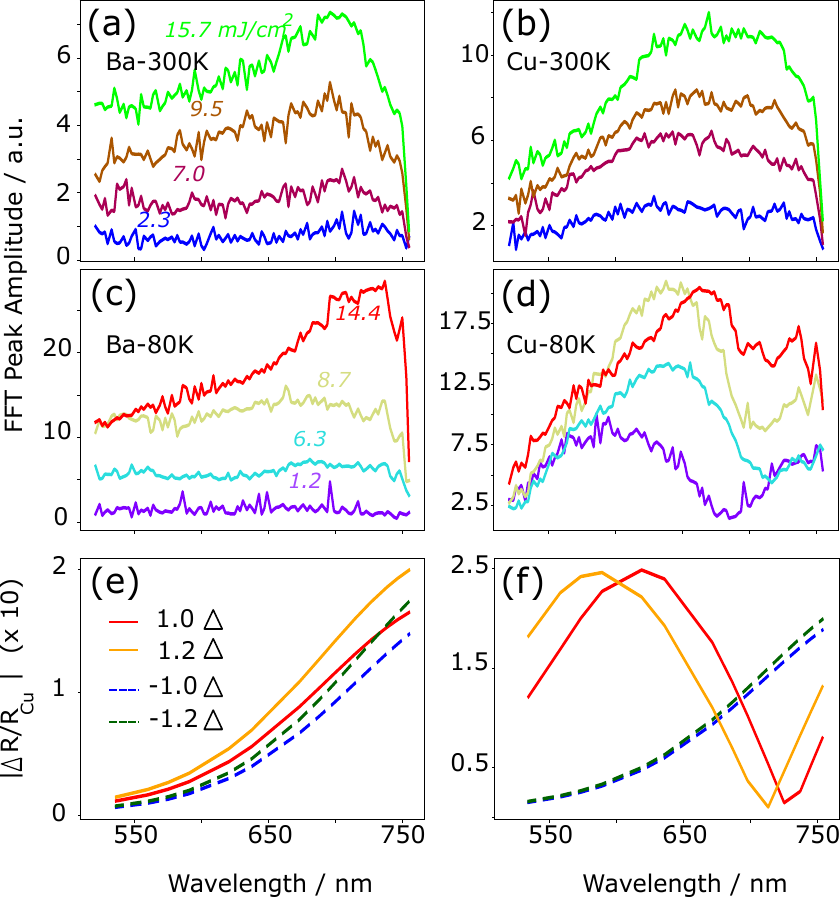}
\caption{\label{ph_opt} Probe wavelengths dependence of the FFT peak amplitude for the barium ((a) 300\,K, (c) 80\,K) and copper ((b) 300\,K, (d) 80\,K) modes.  Simulated reflectivity changes if the coherent phonon modulates the oscillator strength of the Cu-O charge transfer resonance (e), or resonance position (f). In both cases the oscillator strength or resonant energy is increased or decreased by an amount $\Delta$~\cite{SupInfo}. }
\end{figure}

Having established that the Cu mode is anharmonic, we now examine how the phonon can modulate the electronic properties of YBCO. Fig.~\ref{ph_opt} shows the amplitude of the reflectivity change induced by the Cu and Ba modes for several fluences. At room temperature, we find that the shape of the spectral response is independent of excitation for both Ba (Fig.~\ref{ph_opt}(a)) and Cu (Fig. \ref{ph_opt}(b)) modes, in agreement with the linearity of the signal with fluence. At 80\,K, the wavelength dependence of the barium mode (Fig.~\ref{ph_opt}(c)) is broadly wavelength independent for low fluences and starts to increase in amplitude at longer wavelengths for high fluences. However, the copper mode show stronger spectral shifts that depend on pump fluence (Fig.~\ref{ph_opt}(d)). For low fluences, the spectral response is in excellent agreement with previous works and the maximum sensitivity to the displacement is found at $\sim$ 600\,nm \cite{Novelli}. However, higher fluences show a shift in the phonon resonances, with the maximum moving to longer wavelengths, eventually peaking at $\sim$ 670\,nm.

The visible spectral range probed in this experiment is mainly sensitive to electronic copper-oxygen charge transfer resonance, which is situated around 450\,nm~\cite{Romberg}. The position of this resonance is known to anti-correlate with the superconducting transition temperature, i.e. cuprates with the the highest $T_c$ have the charge transfer resonance at higher energy (shorter  wavelength)~\cite{Weber}. Displacing the copper ion should modulate the charge transfer resonance because the Cu-O bond angles and length directly influence the optical transition probabilities between the Cu and O. Therefore, we now investigate whether the spectral dependence of the reflectivity change is consistent with a coherent phonon modulating the position of the transfer resonance. 

In the theory of displacive excitation of coherent phonons~\cite{Zeiger}, photoexcited carriers change the equilibrium position of the phonon coordinate promptly, resulting in oscillations around a new equilibrium position. For a step like force acting on a harmonic potential the time-dependence of the phonon coordinate relative to the initial position is

\begin{equation}
\label{phonon}
\delta Q_a(t)=\frac{Q_a}{2}(1-\cos(\omega_a t)), t >0,
\end{equation}
where $Q_a$ is the amplitude of the displacement induced by photo-excitation and $\omega_a$ is the angular frequency of the specific coherent mode. For an anharmonic potential, additional higher order harmonics will also be generated. In the linear approximation, the change in reflectivity resulting from the phonon ($R_{ph}$) is given by 
\begin{equation}
\label{RoT}
\delta R_{ph}(t)=\frac{\partial R}{\partial \delta Q_a}\delta Q_a  = \frac{\partial R}{\partial \delta Q_a}\frac{Q_a}{2}(1-\cos(\omega_a t)).
\end{equation}

The total transient change in reflectivity is a combination of the electronic ($R_{e}(t)$) and two phonon responses ($R_{a,b}(t)$), 
\begin{equation}
\label{RoTfull}
\begin{split}
\delta R(t) & =R_{a}(t) + R_{b}(t) + R_e(t)\\
&= R_s(t) - R_a\cos(\omega_a t) - R_b\cos(\omega_b t), 
\end{split}
\end{equation}
where $R_s$ contains the non-oscillatory electronic contributions as well as the non-oscillatory parts coming from the phonon displacement and $R_{a,b}=\frac{\partial R}{\partial \delta Q_{a,b}}\frac{Q_{a,b}}{2}$. 

Background subtraction removes the $R_s(t)$ terms, leaving only the oscillatory parts and the Fourier transform of this residual gives $|R_{a,b}|$, which are plotted in Fig.~\ref{ph_opt}a-d. Although the reflectivity change resulting from the quasi-equilibrium shift in the phonon positions are lost in the background subtraction, as the oscillating term is also directly related to the shift it can be unambiguously extracted from the phonon amplitude. Therefore, by examining the spectral dependence of the phonon oscillation, we can examine how the electronic degrees of freedom are modulated by the shift in the phonon coordinate.

In order to model how the optical reflectivity changes due to a modification of the charge transfer resonance position, we calculate the normal incidence reflectivity using the Drude-Lorentz model reported in Ref.~\cite{Romberg}. Figs.~\ref{ph_opt}(e,f) show how the reflectivity changes when the Lorentzian response for the copper-oxygen charge transfer resonance is modified, either through changes in the oscillator strength (e) or through the resonance position (f) by an amount $\pm\Delta$, where $\Delta$ represents the shifted equilibrium position about which the oscillations occur, enabling a direct comparison to spectral response plotted in Figs.~\ref{ph_opt}a-d. Modifying the oscillator strength (e) produces quantitatively similar lineshapes, independent of strength or sign of the shift. A similar feature is observed when the the charge transfer resonance is decreased (f). However, increasing the resonance position produces line-shapes that are similar to what we observe for the Cu mode in Fig.~\ref{ph_opt}(d).

\begin{figure}
    \centering
    \includegraphics{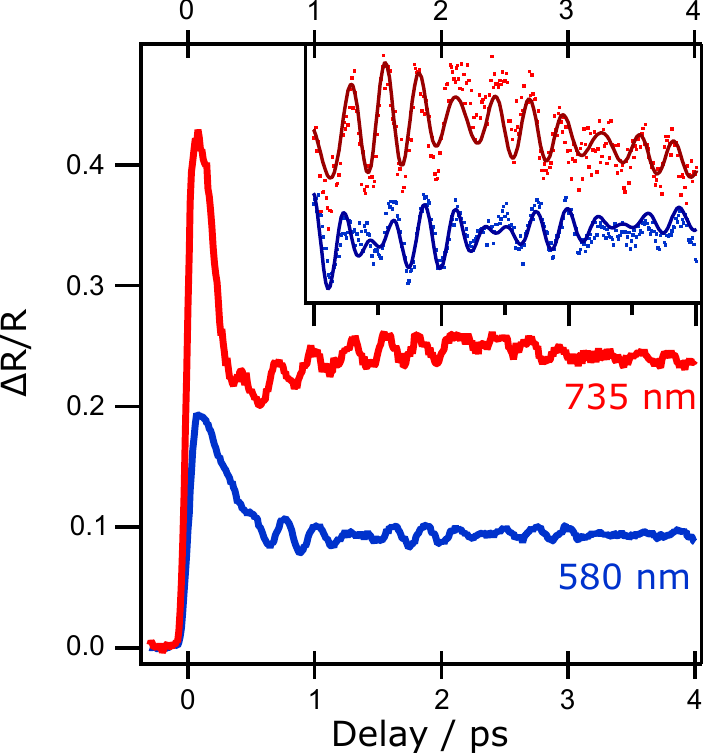}
    \caption{Time-dependent reflectivity at 80 K, and 14.4 mJ/cm$^2$ measured at 580\,nm and 735\,nm. The inset focuses on the oscillating component plotted on the same x-axis as the main graph by offset in y, with the data (square markers) fitted by the sum of two oscillations at 3.5 and 4.5 THz as described in the main text.}
    \label{fig:phase}
\end{figure}

Fig.~\ref{ph_opt} shows the absolute value of the reflectivity change, to enable comparisons to the Fourier transform. However, the change in reflectivity can be either positive or negative and the minima in Fig.~\ref{ph_opt}f corresponds to a change in sign in the reflectivity due to the change in the resonance position, i.e. one side of the charge transfer resonance increases in reflectivity while the other decreases. For an oscillating mode, this corresponds to a $\pi$ phase shift in the coherent phonon response on either side of the spectrum. As the phase information in lost in the Fourier transform, we consult the time-domain. Fig~\ref{fig:phase}, plots two cuts from the data in Fig.~\ref{fig1}b at $\lambda = 580,  735$\,nm. 

Due to the presence of two phonon modes it is not immediately clear how the phase of Cu mode changes. Therefore, to get the phase information we fit the oscillations in the time domain with $A_{Ba}(\lambda) \cos(\omega_{Ba} t) \exp(-t/\tau_{Ba}) + A_{Cu}(\lambda) \cos(\omega_{Cu} t) \exp(-t/\tau_{Cu})$, together with a slowly varying function for the background. $A$, $\omega$ and $\tau$ respectively correspond to the amplitude, angular frequency and decay rate of barium and copper vibrations. We find an excellent fit to the data for cosine-like oscillations, demonstrating that we are in the displacive limit. More importantly, we observe that while $A_{Ba}$ has the same sign for both wavelengths, $A_{Cu}$ changes sign either side of the resonance, demonstrating that the phonon oscillates the resonance position. 

The qualitative spectroscopic agreement suggests that photoexcitation shifts the equilibrium phonon position which in turn shifts the charge transfer resonance to higher energies. In equilibrium, this would correspond to a decrease in the expected superconducting transition temperature. In our case, the shift in the phonon coordinate results from a large excitation of charge which will, in any case, rapidly destroy the superconducting state and may provide additional renormalization of the position of the charge transfer resonance which will dominate any modification of $T_c$ due to the lattice~\cite{Kusar, Giannetti, Coslovich, Rameau}. However, in non-linear phononics~\cite{Forst}, Raman active phonons are displaced through vibrational excitation or IR active modes. This may suppress electronic heating, and could allow the role played by the lattice to dominate. The main issue that needs still to be overcome is the direction of the phonon displacement. In order for $T_c$ to increase, the resonance needs to shift to higher energies, which requires the bond to move in the opposite direction to that induced by light. The current understanding of phonon-phonon coupling is that it is a rectification process similar to optical excitation, which does not allow directional control. Thus it could be expected that IR excitation of the Raman active copper mode will also lead to a decrease in $T_c$ and not an increase. 

\section{Conclusion}

Our results provide a mixed outcome for the origins of light-induced superconductivity. On the one hand, we observe a large anharmonic response of the copper vibrational mode, a key piece of evidence which enables non-linear phonon interactions. In addition, we have shown that the copper mode directly modulates the Cu-O charge transfer resonance, which is known to correlate with $T_c$ in equilibrium. Thus, we provide new evidence that control of $T_c$ through lattice distortions may be possible. The fact that the anharmonic response is lost or suppressed at room temperature may even suggest that this method of coupling becomes less efficient at higher temperatures, and that the loss of anharmonicity may set the temperature scale on which superconductivity can be induced. 

However, on the other hand, to actually enhance $T_c$ requires directional control of the charge transfer resonance. This is currently lacking in the non-linear phononics mechanism. As both optical and phonon excitation generate a displacement through a rectification process, one expects that the direction of the displacement should be the same. Therefore, our results suggest that phonon excitation would actually suppress $T_c$. Understanding how to control the direction of phonon displacements will be key to achieve real control over material structures. Further work is also required to understand how the electronic temperature evolves under photon-excitation, as vibrational modes can still drive significant increases in the electronic temperature, which would again suppress superconductivity~\cite{Murakami}. As a result, experimental approaches that combine optical measurements with time-resolved two photon photoemission~\cite{Yang} and femtosecond X-ray diffraction~\cite{Gerber} are needed in order to fully understand non-equilibrium the processes that occur in phonon driven high temperature superconductors. 

This project has received funding from the European Research Council (ERC) under the European Union's Horizon 2020 research and innovation programme (grant agreement N$^o$ 758461 and was supported by Spanish MINECO (Severo Ochoa grant SEV-2015-0522, SEV-2015-0496), Ram\'on y Cajal programme RYC-2013-14838 and Fondo Europeo de Desarrollo Regional FIS2015-67898-P (MINECO/FEDER), MAT2014-51778-C2-1-R as well as Fundaci\'o Privada Cellex, and CERCA Programme / Generalitat de Catalunya.

\onecolumngrid

\renewcommand{\thefigure}{S\arabic{figure}}
\setcounter{figure}{0}    

\vspace{2cm}
\begin{center}
\textbf{\large Supplemental Material: Probing the Lattice Anharmonicity of Superconducting\\ YBa$_2$Cu$_3$O$_{7-\delta}$ Via Phonon Harmonics}
\end{center}

\vspace{1cm}
{\textbf{Fluence dependence of the third harmonic}}\\

\begin{figure}[b]
\centering
\includegraphics{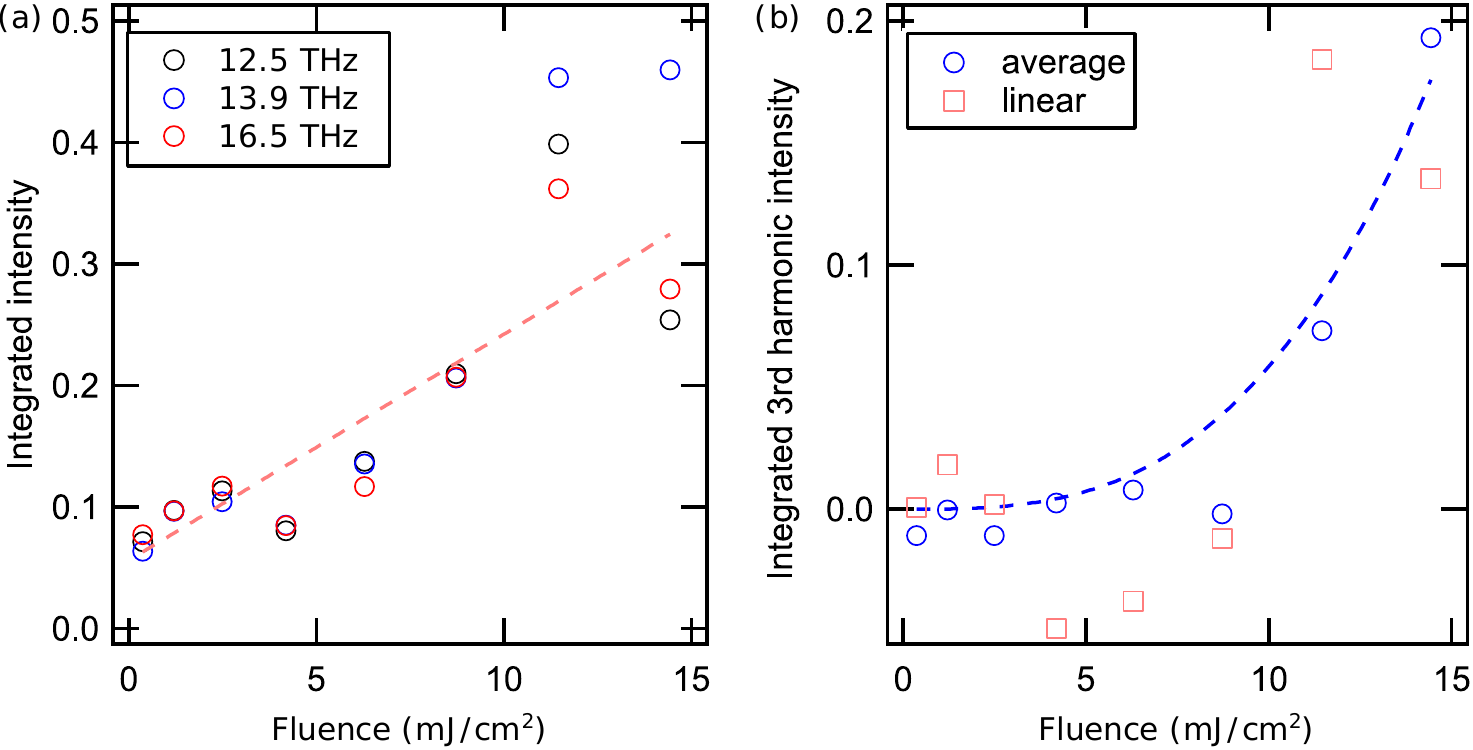}
\caption{\label{Fluence} Dependence of Fourier transform amplitude as a function of fluence in the vicinity of the third harmonic signal. (a) integrated intensity from the Fourier transform shown in Fig 2b of the main manuscript at three different frequencies, together with a linear fit (dashed line). (b) The third harmonic response (13.9 THz) after subtracting the average of the 12.5 and 16.5 THz data (blue circles) and the linear fit (red squares). The dashed blue line is a fit proportional to the cube of the fluence}
\end{figure}

In Fig~\ref{Fluence} we show the fluence dependence of the Fourier intensity around the third harmonic signal. The data presented in Fig 2b is integrated with 1 THz of bandwidth centred at the frequencies specified in the figure legend. There are three primary sources of noise in the data, detector noise, probe intensity fluctuations and pump intensity fluctuations. These fluctuations are uncorrelated with the delay, and thus add a time-independent noise source as a function of the delay, which translates to a white-noise-like frequency response. As the probe and detector do not change for each measurement, the noise from these sources should be constant. In the linear response regime, noise from the pump should scale linearly with fluence. The off-resonant intensity (12.5 and 16.5 THz) can be fit by $I = a + bF$, which demonstrates the above noise dependence (Fig~\ref{Fluence}a). Fig~\ref{Fluence}b shows the third harmonic response after subtracting either the average fluence dependence of the off-resonant signals, or the linear fit to the off resonant signal. The remaining response can be well described by a $I \propto F^3$, as expected for the third harmonic response. 

\vspace{1cm}
{\textbf{Drude-Lorentz model of the Reflectivity}}\\

\begin{figure}
\centering

\includegraphics[scale=1.0]{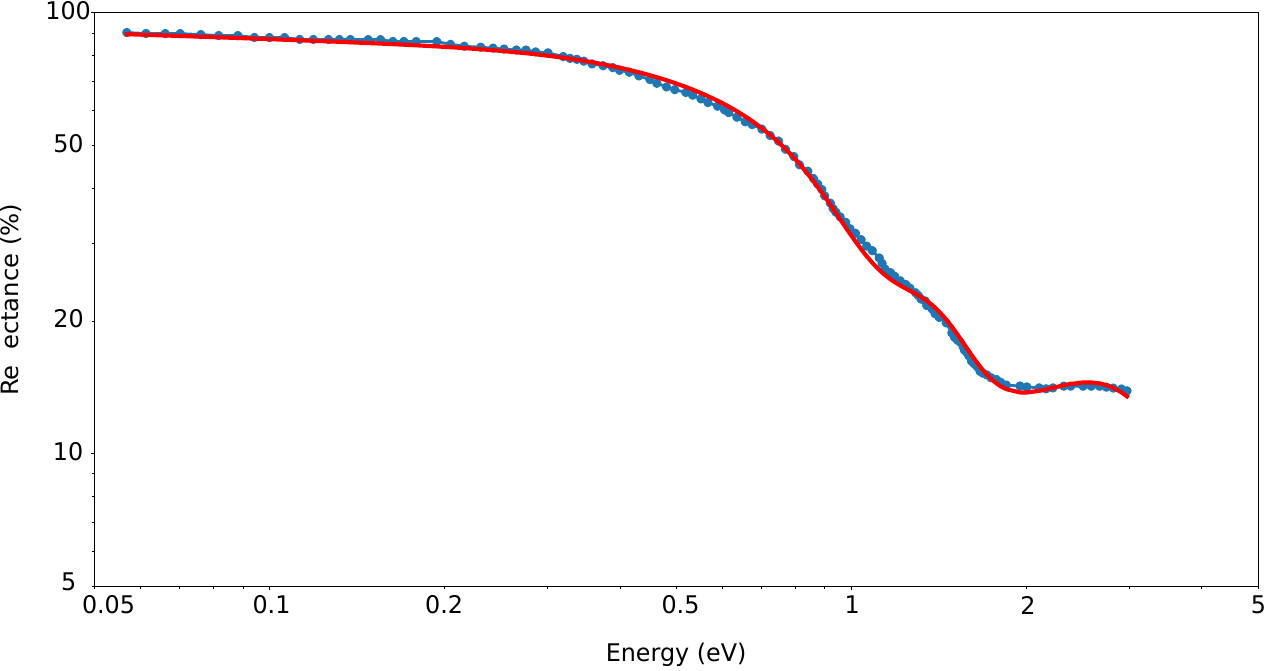}
\caption{\label{fitref} Extracted data from Fig.11 of the Ref. 23 (dots) compared to Drude-Lorentz model fit (solid line)}
\end{figure}

In order to investigate how specific changes to the electronic structure modify the reflectivity, we construct a Drude-Lorentz model for the reflectivity by using the reflectance data at normal incidence captured from Fig.11 of Ref.23 of the main text. The dielectric function was calculated as:
\begin{equation}
\label{epsilon_Drude_Lorentz}
\epsilon(\omega) = \epsilon+ \sum\limits_{i} \frac{A_i}{(\omega_i^2 -\omega^2) + j \gamma_i\omega}
\end{equation}
where $\omega_{i}$, $\gamma_i$ and $A_i$ are the resonance frequency, damping rate and oscillator strength, for the different resonances respectively. These values are taken from Table 1 of Ref. 16 and are not modified for our sample. For simplicity we only include the first four terms and re-scale the amplitude, which already gives a good fit to the data. The resulting reflectivity at normal incidence can be calculated by: 
\begin{equation}
\label{Reflectivity}
R(\omega) = | \frac{1-\sqrt{\epsilon(\omega)} }{1+\sqrt{\epsilon(\omega)} }|^2
\end{equation}
The fit, compared to the original data is shown in Fig~\ref{fitref}.

In the theory of displacive excitation of coherent phonons, the time dependent displacement, relative to the equilibrium condition, for a step like force in a harmonic potential is given by
\begin{equation}
\label{phonon}
\delta Q_a(t)=\frac{Q_a}{2}(1-\cos(\omega_a t)), t >0,
\end{equation}
where $Q_a$ is the amplitude of the displacement induced by photo-excitation and $\omega_a=2\pi f_a$ is the angular frequency of the coherent mode. For an anharmonic potential, additional higher order harmonics will also be generated. For sufficiently small $Q_a$, the change in reflectivity resulting from the phonon ($R_{ph}$) is given by 
\begin{equation}
\label{RoT}
\delta R_{ph}(t)=\frac{\partial R}{\partial \delta Q_a}\delta Q_a  = \frac{\partial R}{\partial \delta Q_a}\frac{Q_a}{2}(1-\cos(\omega_a t)).
\end{equation}

The total transient change in reflectivity is a combination of the electronic ($R_{e}$) and two phonon responses ($R_{a/b}=\frac{\partial R}{\partial \delta Q_{a/b}}\frac{Q_{a/b}}{2}$), 
\begin{equation}
\label{RoTfull}
\delta R(t)=R_{a}(t) + R_{b}(t) + R_e(t) = R_e(t) + R_a + R_b - R_a\cos(\omega_a t) - R_b\cos(\omega_b t). 
\end{equation}
 
Background subtraction removes the $R_e(t)$ terms and the time independent phonon terms, leaving only the oscillatory parts. Fourier transforming this data extracts gives $|R_{a/b}|$ from the oscillatory part. These spectra are plotted in Fig.\,5a-c of the manuscript. The shape of this response can be used to determine how the resonance has shifted due to the phonon displacement. To extract this information, we assume that this phonon displacement only affects the copper-oxygen resonance, either through the modulation of the oscillator strength $A_\mathrm{CuO} \rightarrow A_\mathrm{CuO} \pm \Delta$ or through the resonance condition $\omega_\mathrm{CuO}\rightarrow \omega_\mathrm{CuO} \pm \Delta$, where $\Delta\propto Q(t)$ and the damping rate and other modes in the Drude-Lorentz model are left unperturbed. We then calculate the resulting magnitude of the reflectivity change normalize to the unperturbed value for the two cases, which are plotted in Fig. 5e-f. The best agreement found when $\omega_\mathrm{CuO}$ shifts to higher energies. 

\end{document}